\documentclass[journal]{IEEEtran}

\usepackage{cite}
\ifCLASSINFOpdf
\else
\fi

\usepackage{amsmath}

\ifCLASSOPTIONcompsoc
  \usepackage[caption=false,font=normalsize,labelfont=sf,textfont=sf]{subfig}
\else
  \usepackage[caption=false,font=footnotesize]{subfig}
\fi

\ifCLASSOPTIONcaptionsoff
  \usepackage[nomarkers]{endfloat}
 \let\MYoriglatexcaption\caption
\renewcommand{\caption}[2][\relax]{\MYoriglatexcaption[#2]{#2}}
\fi

\usepackage{tikz}
\usetikzlibrary{shapes.geometric,shapes, arrows,fit, matrix, positioning}
\usetikzlibrary{math}
\usepackage{graphicx}
\usetikzlibrary{arrows.meta}
\graphicspath{{./eps/}}
\usetikzlibrary{calc}
\definecolor{myblue}{RGB}{0, 127, 217}
\definecolor{myred}{RGB}{233, 30, 99}
\definecolor{inclusioncolor}{RGB}{255, 170, 71}
\definecolor{mygrey}{RGB}{127, 127, 127}
\definecolor{mybgd}{RGB}{215, 235, 255}
\definecolor{mygreen}{RGB}{90, 177, 30 }
\usepackage{gensymb}
\usepackage{hyperref}
\usetikzlibrary{matrix}
\usepackage{tabularx}
\usepackage{tabulary}
\usepackage{amssymb}
\usepackage{layouts}
\usepackage{pgfplots}
\usepackage{pdfpages}
\usepackage{transparent}
\newcounter{BlockCounter}
\usepackage{makecell}
\usepackage{subfiles}

\usepackage[capitalise]{cleveref}
\newcommand{\eg}{e.g., }
\newcommand{\ie}{i.e., }

\newcommand{\refBlock}[1]{%
\hyperref[#1]{Block~\ref*{#1}}
}
\newcommand{\rulesep}{\unskip\ \vrule\ }
\usepackage{upgreek}

\begin{document}

\title{Learning the Imaging Model of Speed-of-Sound Reconstruction via a Convolutional Formulation}

\author{
    \IEEEauthorblockN{Can Deniz Bezek\IEEEauthorrefmark{1}, Maxim Haas\IEEEauthorrefmark{2}, Richard Rau\IEEEauthorrefmark{2}, Orcun Goksel\IEEEauthorrefmark{1}\IEEEauthorrefmark{2}} \\[1ex]
    \IEEEauthorblockA{\small\IEEEauthorrefmark{1}Department of Information Technology, Uppsala University, Sweden}\\
    \IEEEauthorblockA{\small\IEEEauthorrefmark{2}Computer-assisted Applications in Medicine Group, ETH Zurich, Switzerland}\vspace{-4ex}
\thanks{Funding was provided by the Uppsala Medtech Science \& Innovation Centre; and for the clinical study partial by the Young Researchers Grant of the European Society of Breast Imaging.}%
\thanks{M.~Haas and R.~Rau contributed the work while they were with the Computer-assisted Applications in Medicine Group, ETH Zurich, Switzerland.}%
\thanks{For the clinical data, the authors would like to thank Dr.\ Rahel A.\ Kubik-Huch and the personnel at the Cantonal Hospital Baden, Switzerland.}%
}

\maketitle

\begin{abstract}
Speed-of-sound (SoS) is an emerging ultrasound contrast modality, where pulse-echo techniques using conventional transducers offer multiple benefits.
For estimating tissue SoS distributions, spatial domain reconstruction from relative speckle shifts between different beamforming sequences is a promising approach.
This operates based on a forward model that relates the sought local values of SoS to observed speckle shifts, for which the associated image reconstruction inverse problem is solved. 
The reconstruction accuracy thus highly depends on the hand-crafted forward imaging model. 
In this work, we propose to learn the SoS imaging model based on data.
We introduce a convolutional formulation of the pulse-echo SoS imaging problem such that the entire field-of-view requires a single unified kernel, the learning of which is then tractable and robust.
We present least-squares estimation of such convolutional kernel, which can further be constrained and regularized for numerical stability.
In experiments, we show that a forward model learned from k-Wave simulations improves the median contrast of SoS reconstructions by 63\%, compared to a conventional hand-crafted line-based wave-path model.
This simulation-learned model generalizes successfully to acquired phantom data, nearly doubling the SoS contrast compared to the conventional hand-crafted alternative.
We demonstrate equipment-specific and small-data regime feasibility by learning a forward model from a single phantom image, where our learned model quadruples the SoS contrast compared to the conventional hand-crafted model.
On in-vivo data, the simulation- and phantom-learned models respectively exhibit impressive 7 and 10 folds contrast improvements over the conventional model.

\end{abstract}


\section{Introduction}
Being low-cost, non-ionizing, and real-time, ultrasound (US) is a widely used medical imaging modality. 
Typical B-mode US images display the reflectivity of tissue structures, which however is only qualitative information.
Biomechanical characteristics, such as shear-modulus that can be estimated via ultrasound elastography, can provide quantitative tissue info.
Speed-of-sound (SoS) is an alternative quantitative biomechanical marker that relates to bulk modulus and may therefore provide complementary and/or independent information on tissue composition and pathological state, \eg for breast cancer~\cite{li_breast_2017,malik_quantitative_2018,malik_breast_2019,ruby_breast_2019}.
Reconstructing local maps of SoS as well as attenuation was proposed using transmission-based US computed tomography~\cite{duric_CURE_2007,gemmeke_3D_2007, li_invivo_2009,mamou_quantitative_2013,li_breast_2017,malik_quantitative_2018,malik_breast_2019}.
This however requires double-sided access to the anatomical structure of interest, \eg using ring shaped~\cite{duric_CURE_2007}, 3D~\cite{gemmeke_3D_2007}, or two opposing~\cite{malik_quantitative_2018} transducer configurations. 
Such systems require bulky and costly setups, which are complicated to operate and are applicable only on body parts that are immersible with access from all sides; thus with main application on the breast.
Alternatively, reflector setups~\cite{krueger_limited_1998,sanabria_reflector_2016,sanabria_SoS_2018} offer a hand-held solution with conventional transducers, allowing for promising use-cases such as breast density estimation~\cite{sanabria_breast_2018} and sarcopenia prediction~\cite{sanabria_sacropenia_2019}.
However, this approach still requires two-sided tissue access, complicating clinical application.

Pulse-echo methods eliminate the above limitations and enable SoS reconstruction using conventional US setups without additional hardware and hence constraints.
For pulse-echo reconstruction of local SoS distribution based on observed speckle-shifts from different plane-wave (PW) transmissions, a Fourier-domain approach was first proposed in~\cite{jaeger_computed_2015}.
A spatial-domain reconstruction method~\cite{sanabria_spatial_18} was shown to yield improved accuracy, and was demonstrated in a clinical application for imaging breast cancer~\cite{ruby_breast_2019}. 
A method for estimating SoS maps by utilizing the spatial coherence of backscattered echoes was introduced in~\cite{imbault_robust_17}, by also incorporating phase aberration correction taking into account fat and muscle layers. 
Adaptive receive aperture~\cite{stahli_improved_20} was proposed to achieve point-spread function invariance.
As a PW alternative, diverging wave transmission~\cite{rau_SoS_2019} was shown to reduce abberation artifacts.
Besides tissue differentiation, local SoS reconstruction can find utility in locally-adaptive beamforming, with promising potential improvements as shown in B-mode images~\cite{rau_ultrasound_2018} and in shear-wave elastography~\cite{Chintada_phase-aberration_21}.

All the above works utilize some form of an assumed forward model that relates medium SoS distribution to observed echo shifts.
Such models are decided and crafted based on the understanding of the involved physical, electrical, and computational phenomena such as sequence timing, wave propagation, and beamforming effects.
Several effects in this pipeline are relatively complex, hence the models often require approximations and simplifications.
For instance, a typical pulse-echo SoS reconstruction forward model considers echo shift sensitivity to SoS only along an infinitesimally-thin geometrical ray path, although to be precise the SoS values within a finite thickness (Fresnel zone) need to be considered given the limited bandwidth nature of transmissions~\cite{Martiartu2019}.
Indeed, for various reasons any nearby structures, \ie SoS variation just outside the considered ray-path line, may affect the ray propagation.
Furthermore, given multi-element apertures, measured time-delays are sensitive to SoS variations not only along a single line, but also those in a larger area within the aperture.
Although forward models can be made more elaborate, several effects are often not straight-forward to accurately take into account, such as complex physical phenomena, unexpected (side)effects of computational operations, discretization, quantization, and other possible systematic errors and offsets.

To be able to incorporate difficult-to-model phenomena and potentially unknown contributors in a forward model, in this work we propose a novel approach for learning an SoS reconstruction forward model from data.
To that end, we propose a convolutional formulation of the SoS imaging problem, which helps to reduce the number of unknowns to be learned to a single unified convolutional kernel (instead of the exponentially larger full forward-model matrix).
This allows for easily introducing constraints and regularization, enabling learning from a few and even a single sample image, with greatly improved numerical stability and resulting reconstructions.

\markboth{Bezek \MakeLowercase{\textit{et al.}}: Learning the SoS imaging model}
{Bezek \MakeLowercase{\textit{et al.}}: Learning the SoS imaging model}

\section{Methods}
\subsection{Pulse-Echo SoS Imaging from Speckle Shifts}
Pulse-Echo SoS imaging based on speckle shifts is a limited-angle computed tomography problem, with algebraic solution formulations~\cite{sanabria_spatial_18,rau_SoS_2019}. 
In this approach, the echo shifts observed from different acoustic paths are related to the spatial SoS distribution through a sensitivity matrix representing the forward model of SoS imaging.
To reconstruct SoS distributions, the tissue is insonified multiple times with ultrasound waves, \eg with plane-wave (PW) transmission (Tx), and corresponding images are beamformed from the acquired raw data using an assumed tissue SoS value of~$c_0$.
Speckle shifts (apparent motion / displacement) are then found between a pair of beamformed frames.

Let $\boldsymbol{d}_p$$\in$$\mathbb{R}^{N_x\times N_z}$ be the axial displacement measurements for a frame pair $p$, with $\boldsymbol{t}_p$=$\boldsymbol{d}_p/c_0$ being its time-delay equivalent. 
For the sought SoS map $\boldsymbol{c}$$\in$$\mathbb{R}^{N_x'\times N_z'}$, let $\boldsymbol{\sigma}=1/\boldsymbol{c}$ be the corresponding slowness map and $\boldsymbol{s}=\boldsymbol{\sigma}-\sigma_0$ be the differential slowness map relative to beamforming slowness $\sigma_0$.
Then, for pair $p$ the sensitivity of time-delays to relative slowness distribution, \ie the imaging model, can be encoded as a (sparse) linear matrix $L_p$$\in$$\mathbb{R}^{N_xN_z\times N_x'N_z'}$, which leads to the pulse-echo SoS forward problem $L_p\underline{\boldsymbol{s}}=\underline{\boldsymbol{t}}_p$.
Hereafter, underlined variables denote vectorized forms of 2D spatial maps.

\begin{figure}
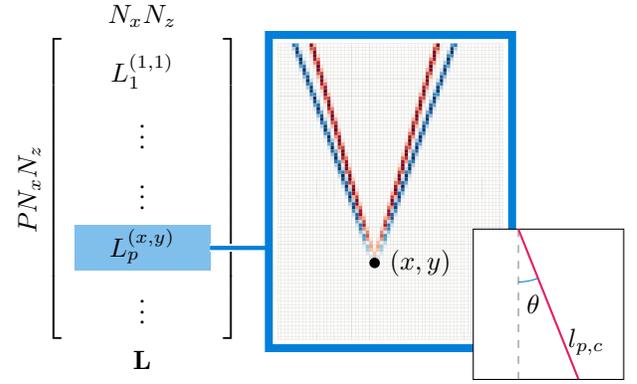

    \centering
    \include{tikz/paths} \vspace{-5ex}
        \caption{Illustration of the differential path matrix $L$, which is a concatenation of $L_p$ for image pairs $p=1..P$.
    In a conventional hand-crafted forward model, each row $L_p^{(x,y)}$ contains the vectorized form of discretized ray paths to an image point $(x,y)$ between pair $p$. 
    Each matrix element thus contains the length of ray falling in the corresponding Cartesian cell.
    Each row then contains positive (red) and negative (blue) paths, one each for transmit and receive, along the PW steering angles ($\theta_{p}^1$, $\theta_p^2$) with respect to $(x,y)$.}
\label{fig:Lmatrix}
\end{figure}

Commonly, models and measurements from multiple image pairs are combined, to fully cover the spatial grid, to better condition the problem, and for robustness against noise. 
With regularization, this yields the following SoS inverse problem with a total-variation formulation for finding an optimal relative slowness map~\cite{rau_SoS_2019}:
\begin{IEEEeqnarray}{rCl}
    \underline{\boldsymbol{\hat{s}}} = \arg\min_{\underline{\boldsymbol{s}}} \|L\underline{\boldsymbol{s}} - \underline{\boldsymbol{t}}\|_1 + \lambda \|D\underline{\boldsymbol{s}}\|_1\,,
    \label{eq:SoS_recon}
\end{IEEEeqnarray}
where $D$ is a regularization matrix, $\lambda$ is the regularization weight, $\boldsymbol{\underline{s}}$$\in$$\mathbb{R}^{N_x'N_z'}$ is the vectorized relative slowness map sought, $\boldsymbol{\underline{t}}$$\in$$\mathbb{R}^{PN_xN_z}$ is the vectorized time-delay measurements concatenated from $P$ pairs, and $L=[L_1^T \cdots L_P^T]^T\in\mathbb{R}^{P N_x N_z\times N_x' N_z'}$ is the concatenated full model matrix with time-delay sensitivity of each grid point along differential paths, as illustrated in~\cref{fig:Lmatrix}.
With $\underline{\boldsymbol{\hat{s}}}$ optimized above, the SoS distribution is then recovered as $\boldsymbol{\hat{c}} = 1/(\boldsymbol{\hat{s}} + \sigma_0)$\,.

\subsection{Estimating the SoS Imaging Model from Data}

The above inverse problem aims to find SoS map, given a known forward model.
Conversely, given that the SoS distribution is known for a time-delay measurement, the imaging model matrix elements may be found using
\begin{IEEEeqnarray}{rCl}
    \hat{L} = \arg\min_{L} \|L\boldsymbol{\underline{s}} - \boldsymbol{\underline{t}}\|_2^2\ .
    \label{eq:L_classical_formulation}
\end{IEEEeqnarray}
However, considering no structure in $L$ and in the problem formulation, this would be a highly underdetermined problem due to many degrees of freedom in $L$.
Even with sufficiently many measurements available, a solution is rather intractable (potentially also an ill-posed problem).
To learn the imaging model from data, we thus seek for a model structure that would effectively constrain the problem.

\subsection{Convolutional Formulation of Spatial SoS Reconstruction}
Note that the image point $(x,z)$ and the PW steering angle pair $(\theta_{p}^1$, $\theta_p^2)$ are sufficient to uniquely determine a path matrix row $L_p^{(x,z)}$, cf.~\cref{fig:Lmatrix}.
For a pair $p$, since the PW steering angles to every echo shift measurement location are constant, the 2D grid representations of different model rows are then simple translations of each other.
Formally, row $L_p^{(x+\Delta x,z+\Delta z)}$ seen as a 2D indexed image is the $(\Delta x,\Delta z)$ shifted version of row $L_p^{(x,z)}$.
This observation hints at the above SoS forward model having a convolutional form.

\begin{figure}
    \centering
        \subfloat[Convolutional Model]{{\includegraphics[width=0.75\linewidth]{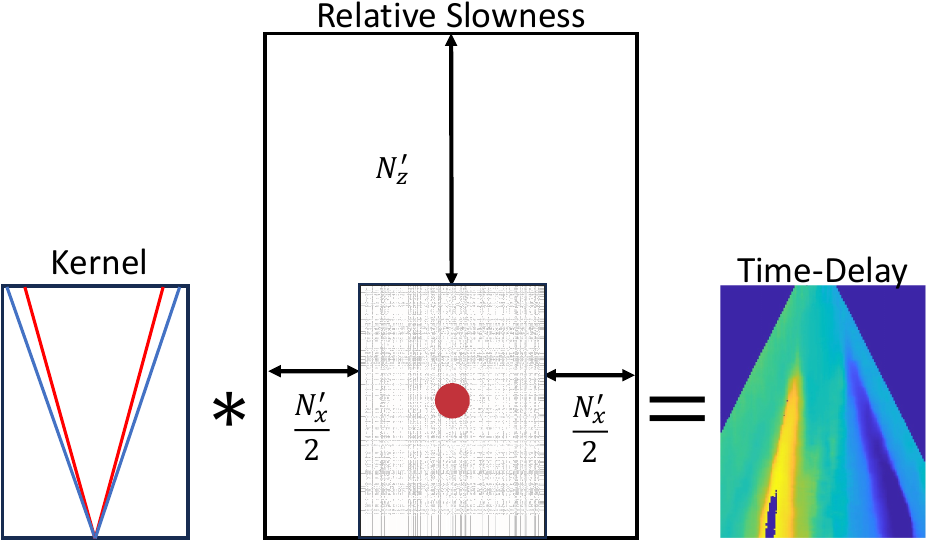}}
     \label{fig:convolution_model}}
     \\
    \subfloat[Sliding Convolutional Kernel]{{\includegraphics[width=0.75\linewidth]{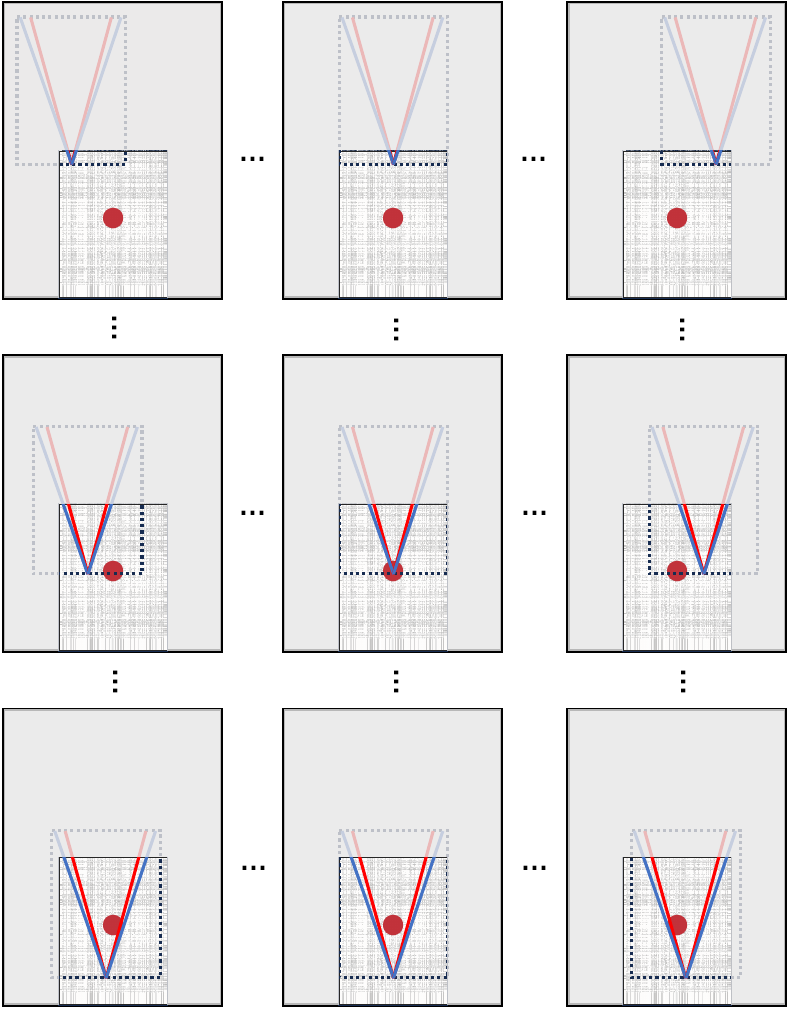}}
\label{fig:sliding_convolution_multiple}}  
    \caption{(a) Illustration of our proposed convolutional model.
    (b) Illustration of a fixed kernel sliding over a zero-padded (gray-area) relative slowness map example, where the map overlapped area approximates kernels for the corresponding imaged points}
    \label{fig:summary}
\end{figure}

To be able to write the problem convolutionally, we define kernel $\boldsymbol{k}_p$ as the differential path operator for the mid-bottom (deepest center) image point $(N_x'/2, N_z')$, in order to convolve this kernel with the relative slowness map.
The kernel choice allows for maximum possible non-zero entries of all paths, \ie the longest possible paths with the shorter forms thus being image crops thereof.
To utilize this kernel at shallower image points, parts of the kernel that fall outside the field-of-view can be ignored (nullified).
For this, we define $\boldsymbol{\tilde{s}}$ by zero padding map $\boldsymbol{s}$ by kernel height on top and half-kernel-width on both sides, as seen in~\cref{fig:convolution_model}.
A convolutional forward imaging problem for the steering angle pair $p$ can then be written as
\begin{IEEEeqnarray}{rCl}
   \boldsymbol{k}_p \ast \boldsymbol{\tilde{s}} =  \boldsymbol{t}_p\,.
    \label{eq:L_convolutional_k}
\end{IEEEeqnarray} 
The sliding of a sample kernel over the padded slowness map is illustrated in~\cref{fig:sliding_convolution_multiple}.

\subsection{Learning Convolutional Imaging Kernels}
Learning such kernels require substantially less degrees-of-freedom ($P N_x' N_z'$ unknowns for $P$ Tx pairs), compared to learning the entire matrix $L$ with $P N_x N_z N_x' N_z'$ elements, while the kernels would still allow us to effectively deduce the entire SoS imaging model.
Using the convolutional model~\eqref{eq:L_convolutional_k}, an optimal kernel for pair $p$ is learned via least-squares fitting
\begin{IEEEeqnarray}{rCl}\label{eq:kernel_estimation_generic}
   \underline{\boldsymbol{\hat{k}}}_p = \arg\min_{\underline{\boldsymbol{k}}_p} \| \underline{\boldsymbol{k}_p \ast \boldsymbol{\tilde{s}}} - \underline{\boldsymbol{t}}_p||_2^2\,.
\end{IEEEeqnarray}
Rewriting the convolution in its matrix-vector product form
\begin{IEEEeqnarray}{rCl}
   \underline{\boldsymbol{k}_p \ast \boldsymbol{\tilde{s}}}&=&S \underline{\boldsymbol{k}}_p\,,
    \label{eq:convolution_matrix_transition}
\end{IEEEeqnarray} 
where $S \in\mathbb{R}^{N_x N_z\times N_x' N_z'}$ is the padded relative slowness map written as Toeplitz matrix.
Given a known (groundtruth or assumed) slowness map, the above formulation \eqref{eq:kernel_estimation_generic} then has a closed form solution for pair $p$ as follows:
\begin{IEEEeqnarray}{rCl}
     \underline{\boldsymbol{\hat{k}}}_p &=& \arg\min_{\underline{\boldsymbol{k}}_p} \|S\underline{\boldsymbol{k}}_p - \underline{\boldsymbol{t}}_p\|_2^2
   +\lambda_k \|D_k\underline{\boldsymbol{k}}_p\|_2^2\\
   &=& (S^\text{T}S + \lambda_k  D_k^\text{T} D_k)^{-1}{S}^\text{T}\underline{\boldsymbol{t}}_p\,,
   \label{eq:L_uncontrained_with_reg}
\end{IEEEeqnarray}
where $D_k$ is a kernel regularization matrix penalizing axial and lateral derivatives with weight $\lambda_k$.
Hereafter, this approach is referred as \textit{unconstrained}.

For a robust estimate, given multiple samples (time-delay ground-truth pairs) for learning, we solve the kernel jointly across these samples, by stacking the measurements and corresponding padded slowness matrices as
\begin{IEEEeqnarray}{rCl}
\mathbb{T}_p = 
\begin{bmatrix}
        \underline{\boldsymbol{t}}^1_p\\
        \underline{\boldsymbol{t}}^2_p\\
        \vdots\\
        \underline{\boldsymbol{t}}^{N_S}_p\\
\end{bmatrix},\quad
\mathbb{S} = 
\begin{bmatrix}
        {S}^1\\
        {S}^2\\
        \vdots\\
        {S}^{N_S}\\
\end{bmatrix}\
\label{eq:overall_delta_tau_s_k}
\end{IEEEeqnarray}
such that the sought joint kernel across $N_S$ samples is then
\begin{IEEEeqnarray}{rCl}
  \underline{\boldsymbol{\hat{k}}}_p &=& (\mathbb{S}^\text{T} \mathbb{S} + \lambda  D_k^\text{T} D_k)^{-1} \mathbb{S}^\text{T} \mathbb{T}_p\,.
 \label{eq:widening_kernels_final}
\end{IEEEeqnarray}
\subsection{Constraining the Learned Kernel Space}
Despite being regularized, the solutions with the unconstrained approach above admits arbitrary propagation paths, some of which may be far from being physically plausible, especially with noisy and small learning datasets. 
Nevertheless, we can inject prior information, \eg about the geometric form of plausible acoustic paths, in order to 
constrain the kernel estimation, thereby restricting its solution space.

The sought kernel is the SoS sensitivity map of the mid-bottom displacement tracking point for a given Tx.
Neglecting secondary reflections, etc, one can safely assume the general overall propagation region of the wave front \emph{path} to that point, \eg within 5-10\,mm precision.
However, the exact path and sensitivity \emph{profile} within this region is very difficult to know.
To leverage this as a prior, instead of learning the kernel with independent elements, we rather assume the viable beam path as given and learn its lateral widening profile.
Such widening can, for example, account for the Fresnel zone around the wave path or the lateral SoS averaging due to the readings across the aperture being summed in delay-and-sum.
These require the kernel to encode SoS sensitivities not only at a single pixel but instead within a wider window laterally.
Although one can attempt to theoretically model, \ie hand-craft, such widening function (for which we show a comparative example in our results), with a similar argument to kernels it is indeed challenging to include all possible effects into such widening model, for which a learning based approach is quite advantageous.

We redefine the sought kernel $\boldsymbol{k}_p$ as a combination of path $\boldsymbol{g}_p$ and profile $\boldsymbol{f}_p$.
For the assumed path, without loss of generality, we utilize the earlier centerline approximation shown in \cref{fig:convolution_model}.
We define the beam profile $\boldsymbol{f}_p$ as a set of 1D smoothing filters in the lateral direction that widens and shapes the centerline as needed to take into account varying SoS sensitivity profile along the main path.
Stacking the lateral profiles (\ie smoothing filters) of fixed maximum length $N_c$ at each depth $N_z'$ separately for Tx and Rx halves of the path, we can define a profile matrix $\boldsymbol{f}_p\in\mathbb{R}^{2 N_z' \times N_c}$.
Accordingly, the kernel can be reconstructed by (lateral) 1D-convolution of the path centerline with the corresponding profile at each path distance (depth) as
\begin{IEEEeqnarray}{rCl}
\boldsymbol{k}_p &=&   \boldsymbol{f}_p\,\tilde{\ast} \,\boldsymbol{g}_p
\label{eq:estimation_split}
\end{IEEEeqnarray}
with $\tilde{\ast}$ representing lateral convolution between matrix rows, \ie every row of the left side being convolved only with the corresponding row of the right side.
We then learn the varying lateral window function, \ie the profile matrix $\boldsymbol{f}_p$.

Using the same solution strategy as before, we rewrite the above horizontal convolution in a matrix-vector multiplication form $\underline{\boldsymbol{k}}_p = {G}_p\underline{\boldsymbol{f}}_p$, where $\underline{\boldsymbol{f}}_p \in\mathbb{R}^{2 N_c N_z'\times 1}$ is the vectorized beam profile and ${G}_p \in\mathbb{R}^{N_x' N_z'\times 2 N_c N_z'}$ is 
the paths $\boldsymbol{g}_p$ in a corresponding Toeplizt matrix form considering the horizontal convolutions.
The following optimization problem is then formulated, with the corresponding closed-form solution for finding the beam profile
\begin{IEEEeqnarray}{rCl}
  \underline{\boldsymbol{\hat{f}}}_p &=& \arg\min_{\underline{\boldsymbol{f}}_p}\|S{G}_p\underline{\boldsymbol{f}}_p - \underline{\boldsymbol{t}}_p\|_2^2
   +\lambda_f \|D_f\underline{\boldsymbol{f}}_p\|_2^2\\
 &=& ({G}_p^\text{T} {S}^\text{T} S{G}_p + \lambda_f  D_f^\text{T} D_f)^{-1}{G}_p^\text{T}{S}^\text{T}\underline{\boldsymbol{t}}_p\,,
 \label{eq:Lhat_constrained}
\end{IEEEeqnarray}
where the weight $\lambda_f$ and matrix $D_f$ is for regularization defined as follows:
We penalize the lateral derivative to ensure relatively smooth window functions (profile along the beam width) and we penalize the axial derivative to discourage abrupt variations of such window function along the wave propagation path, \ie the function profiles at two consecutive depths not to differ substantially.
We also include rows in $D_f$ to force the left and right sides of $\boldsymbol{f}_p$ to zero, to ensure that the found window functions are tapered to prevent discontinuity.

If multiple training samples are available to learn the imaging model, similarly to its unconstrained counterpart, the variables $S$ and $\boldsymbol{t}_p$ in~\eqref{eq:Lhat_constrained} are replaced with their stacked versions in~\eqref{eq:overall_delta_tau_s_k} by concatenating the groundtruth SoS maps and the corresponding time-delay observations from the available training samples.
This allows for jointly solving~\eqref{eq:Lhat_constrained} with all the samples, for a robust estimate of profile $\underline{\boldsymbol{\hat{f}}}_p$.

The learned kernel for Tx pair $p$ can subsequently be recovered by
\begin{IEEEeqnarray}{rCl}
    \underline{\boldsymbol{\hat{k}}}_p &=&  {G}_{p} \, \underline{\boldsymbol{\hat{f}}}_p\,. 
    \label{eq:L_constrained_estimation}
\end{IEEEeqnarray} 
This approach is called \emph{constrained} in our results, and used as our default kernel learning method, unless stated otherwise.

\section{Experiments and Implementation Details}
\subsection{Compared Methods}
We compare the kernels learned using our unconstrained and constrained methods, with two other baseline approaches.
The first is the conventional approach from the literature~\cite{sanabria_spatial_18, rau_SoS_2019,schweizer_robust_23}, where the imaging model is formed by assuming the wave propagation (\ie time-delay sensitivity to SoS) along a thin line that connects the time-delay estimation grid point to the desired aperture centre on the RX path and the hypothetical wave origin for that grid point on the TX path.
For a PW, discretizing this line on the SoS map then looks as in \cref{fig:Lmatrix}.
We call this model \textit{hand-crafted line} in our results.

As mentioned earlier, the conventional approach above ignores potential sensitivity of speckle-shifts to SoS values outside a single line, \eg due to speckle shifts being potentially affected from the signal at multiple elements that coherently combine during beamforming.
To isolate the contributions of simply having a widening window function, from our proposed learning of such functions, we introduce a second baseline using a \emph{hand-crafted window}.
To that end, we empirically tried various approaches, \eg using different functions of apodization weights, tapering these towards the tracked point, etc.
A depth-dependent Hann window worked best, potentially due to its mimicking of coherent summation in beamforming.
The profile of this windowing baseline is shown later in our results.

Note that the above is one example from arbitrarily many profile alternatives that one can hand-craft to purportedly model different imaging phenomena.
Our proposed learning approach instead aims to estimate such window profile and hence the imaging model from data, to eliminate the need for manual design and tuning.

\subsection{Datasets}
We conducted experiments both with numerical and tissue-mimicking phantoms.
Numerical phantoms were simulated using the k-Wave toolbox~\cite{Kwave} on a spatial grid of 40$\times$55\,mm, based on a computational model of the linear-array transducer used in the experiments described later.
The spatial and temporal simulation resolutions were set to 75\,$\mu$m and 6.25\,ns, respectively.
We created two simulation datasets: 

\vspace{1ex}\noindent{\bf Blob set}
contains numerical phantoms with inclusions generated by random deformations of random ellipses, as in~\cite{vishnevskiy_image_2018}.
Background SoS value is uniformly sampled from $[1470, 1550]$\,m/s, either being constant or spatially varying around the sampled SoS value.
Inclusion SoS is set to higher or lower than the background SoS, with a maximum SoS contrast of $10$\,m/s.
We generated 96 numerical phantoms, divided randomly into three subsets of 32 samples: training, validation, and test.
The training set was used solely to learn the kernels for the forward model.
The validation set was used to finetune the parameters for kernel learning and local SoS reconstructions. 
The test set was then used to evaluate the performance of learned kernels, on unseen data.

\vspace{1ex}\noindent{\bf Geometric set}
is the test set from~\cite{melanie_training_2020} and contains 28 circular and 4 rectangular inclusions of varying sizes, locations, and SoS contrasts (up to 100\,m/s), on constant or spatially varying background SoS. 
This set allows us to assess learned kernels on out-of-domain data coming from a different distribution.

\vspace{1ex}\noindent{\bf Phantom data}
is from a custom speed-of-sound phantom (CIRS, Norfolk, VA, USA) with a background SoS value of 1509\,m/s and sphere inclusions with manufacturer-reported SoS values of 1588, 1542, 1515, and 1447\,m/s.
Hereafter, these inclusions are called respectively \emph{Inc\,I}, \emph{II}, \emph{III}, and \emph{IV} for convenience. 
We collected the phantom data with UF-760AG system (Fukuda Denshi, Tokyo, Japan), using a linear-array transducer FUT-LA385-12P with 128 channels and $300\,\mu$m pitch. 
We streamed the full-matrix raw RF channel data over a high bandwidth link and stored it on a dedicated PC for processing retrospectively, by 4$\times$ upsampling, beamforming, displacement tracking, kernel learning, and subsequent local SoS reconstruction.

\vspace{1ex}\noindent{\bf Clinical data} 
was collected in-vivo from breast lesions in a clinical study conducted at the Cantonal Hospital Baden, Switzerland, with ethics approval and informed consent (EKNZ, Switzerland, BASEC 2020-01962). 
In this study, the same machine, transducer, and processing pipeline as above were used.
During data collection, the sonographer used B-mode images for probe navigation.
When a location with a suspicious lesion was found, the SoS acquisition sequence was initiated while maintaining the transducer as steady as possible to avoid possible confounding artifacts from motion~\cite{schweizer_robust_23}.

\subsection{Evaluation Metrics}
\label{sec:evaluation_metric}

We use root mean square error (RMSE) to evaluate the accuracy of the reconstructed spatial SoS maps as
 \begin{IEEEeqnarray}{rCl}
 \label{eq:RMSE_SoS}
\mathrm{RMSE}_c= \sqrt{\frac{1}{N_x'N_z'}\sum_{m=1}^{N_x'N_z'}\left(\mathbf{c}_m -\mathbf{{c}}_m^\ast\right)^2}\,, 
 \end{IEEEeqnarray}
where $\mathbf{c}$ and $\mathbf{c}^\ast$ are respectively the ground truth and reconstructed SoS maps, with $m$ indexing their pixels.

We assess the forward model accuracy by comparing the measured and predicted time-delays using RMSE as 
\begin{IEEEeqnarray}{rCl}
\label{eq:DTRMSE}
\mathrm{RMSE}_t= \sqrt{\frac{1}{P N_x N_z}\sum_{n=1}^{P N_x N_z} \left( L \underline{\boldsymbol{s}} - \underline{\boldsymbol{t}}\right)^2}\,,
 \end{IEEEeqnarray}
where $P N_x N_z$ is the total number of time-delay measurements after near-field/edge masking and cross-correlation based thresholding.
Note that this assesses the fit of the SoS forward model to the readings (first term in \eqref{eq:SoS_recon}) in $\ell^2$ space.

SoS reconstruction contrast for inclusions are reported using absolute SoS difference
\begin{IEEEeqnarray}{rCl}
\text{$\Delta$SoS}= |\tilde{\mu}_\mathrm{inc} -\tilde{\mu}_\mathrm{bkg}|     \end{IEEEeqnarray}
between the inclusion (inc) and background (bkg), where $\tilde{\mu}$ denotes the median SoS value in the respective region.
For the numerical phantoms, we use the known inclusion locations from ground-truth SoS maps. 
For the tissue-mimicking phantoms and in-vivo data, we delineate the inclusions on B-mode images, and as the background we use a 5\,mm margin around the inclusion.

\subsection{Implementation Details}
Both for numerical simulations, phantom, and clinical data, we acquired full-matrix raw RF data using single-element (diverging wave) transmissions.
We used a Tx pulse of 4 half cycles and 5\,MHz center frequency, and an Rx sampling frequency of 40.96\,MHz. 
Using the full-matrix capture echo data above, we synthetically simulated plane-wave Tx data for various steering angles retrospectively. 
A dynamic and adaptive receive aperture~\cite{stahli_improved_20} aligned at 0$\degree$ was used with an f-number of 1. 
For displacement tracking we used a cross-correlation-based block matching method~\cite{reza_disp} and utilized only the measurements that are above a predetermined cross-correlation threshold. 
For local SoS reconstruction, the tracked displacements in the near-field and around frame edges were masked out as they often contain large errors.
Kernel learning did not require any measurement masking since the constraints and regularization effectively filtered out errors.

For reconstructions, we used plane-wave Tx pairs \mbox{\{(-20,-16),}\,(-15,-11),\,...\,,\,(10,14),\,(15,19)\}$\degree$, \ie with 4$\degree$ angular disparity and $5\degree$ pair increments, which were found empirically on the validation set. 
We then determined the cross-correlation threshold by minimizing the reconstruction RMSE on the validation set. 
We set the filter length $N_c$ to be 21 reconstruction grid pixels, and defined the regularization weights via visual observations, again on the validation set.

\section{Results}
\subsection{Learned Imaging Models}
\begin{figure}
\centering
\includegraphics[width=\linewidth]{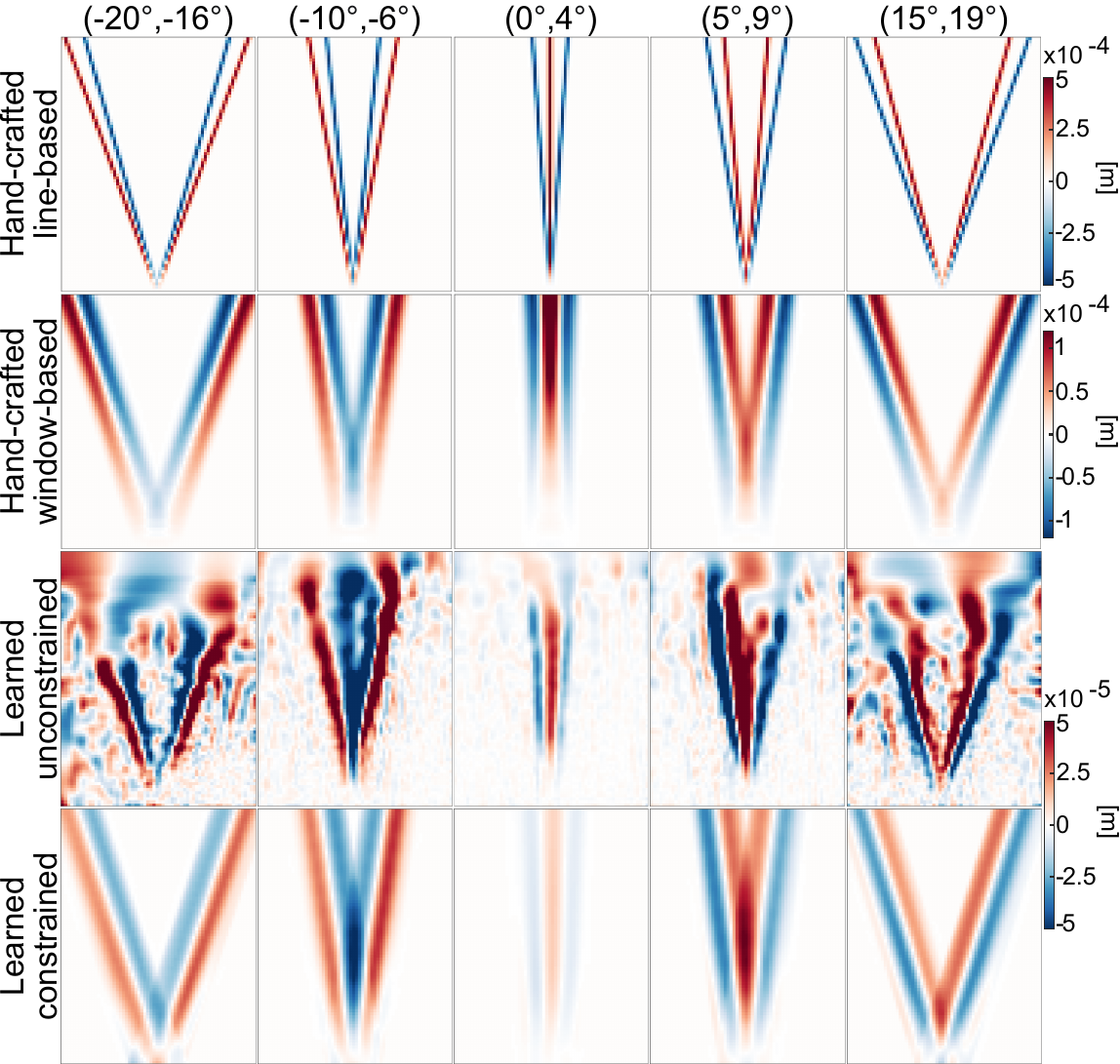}
\caption{Kernels learned using the proposed unconstrained and constrained methods as well as the hand-crated line and window based models, shown for various steering angle combinations. 
For the hand-crafted models, the differential path matrices for the mid-bottom image point are shown for comparability with the learned kernels.
We use different colorbars due to the different scales of various approaches. }
\label{fig:unconstrained_vs_constrained}
\end{figure}

With the training subset of Blob set, we learned kernels using our unconstrained and constrained approaches, to compare these with hand-crafted models with lines and windows.
\Cref{fig:unconstrained_vs_constrained} demonstrates these for various Tx pairs. 
The unconstrained approach yields a noisy kernel, with unrealistic nonzero values (\eg near bottom corners) through which the beam arriving at the mid-bottom point can naturally not pass under reasonable expectation. 
Indeed, the reconstructions using the unconstrained models typically fail due to kernels being unstable (and potentially overfitted to the training data).
The proposed constrained approach yields artifact-free smooth kernels that respect physical expectations and to some extent also resemble the hand-crafted windows. 
Accordingly, for our subsequent reconstruction results we evaluate only the proposed constrained learning approach.
Note that the constrained approach also offers a computational advantage over its unconstrained counterpart, since the former solves for reduced number of unknowns.

\subsection{SoS Reconstructions: Numerical Simulations}
\begin{figure*}
\vspace*{\fill}
    \centering          
    \subfloat[Blob set]{
         \includegraphics[width = .485\linewidth]{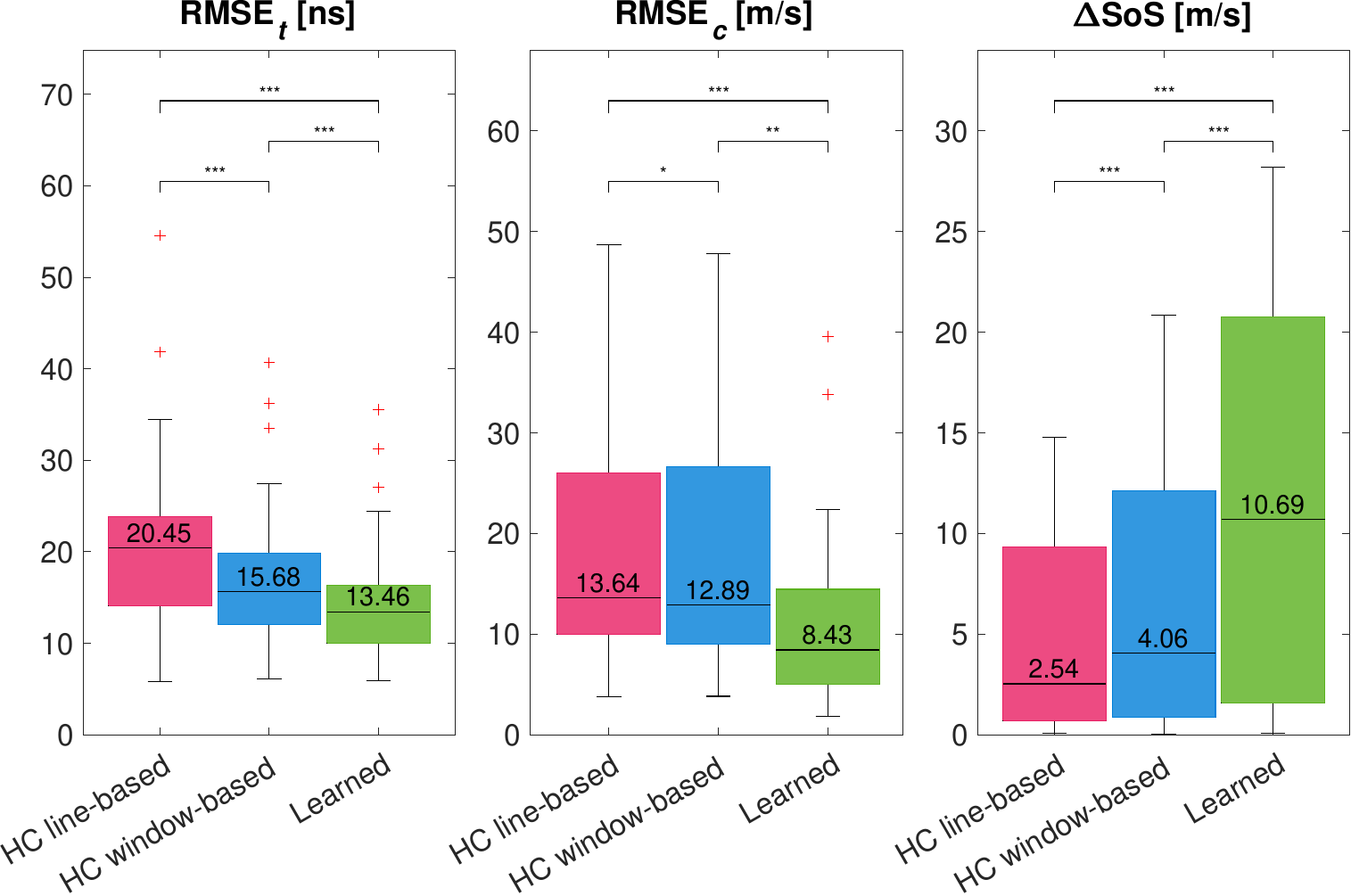}
            \label{boxplot_PW_blob}}
            \hspace{-0.06in}
            \rulesep
            \hspace{-0.11in}
 \subfloat[Geometric set]{
         \includegraphics[width = .485\linewidth]{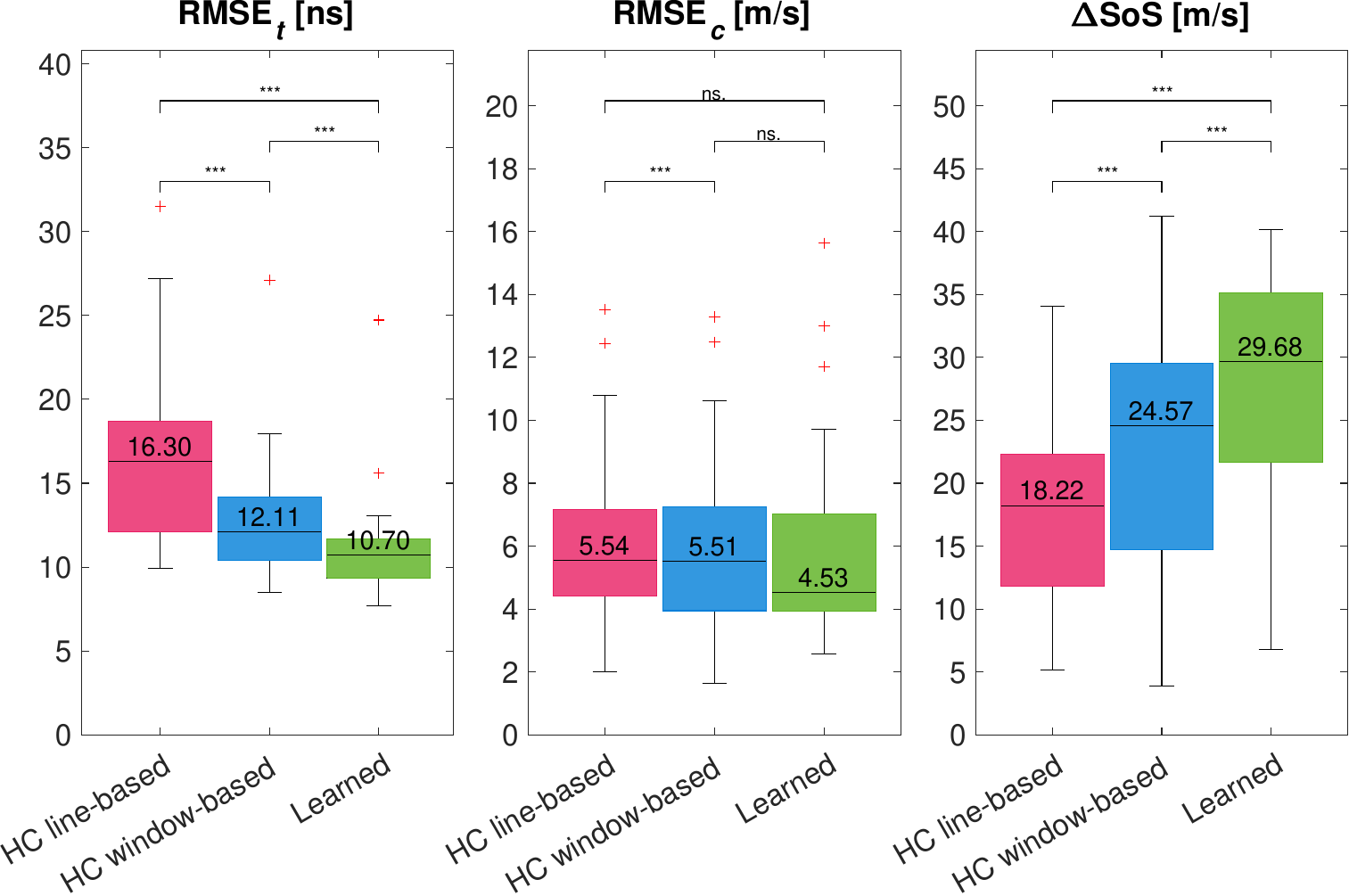}
            \label{boxplot_PW_circle}}    
 \caption{Comparison of models with the hand-crafted (HC) line and windows to our learned model, using forward model accuracy (RMSE$_t$), SoS reconstruction accuracy (RMSE$_c$), and SoS contrast ($\Delta$SoS). Distributions are shown over (a)~the Blob set test samples and (b)~the Geometric set samples, where statistical significance using paired Wilcoxon test are reported using ``***'' for p-value$\leq$0.001, with ``**'' for 0.001$<$p-value$\leq$0.01, with ``*'' for 0.01$<$p-value$\leq$0.05, and ``ns'' for no statistical significance with a higher p-value.
}
\label{fig:DTRMSE_RMSE_DeltaSoS}
\end{figure*}

\begin{figure*}
\centering
\includegraphics[width=\linewidth]{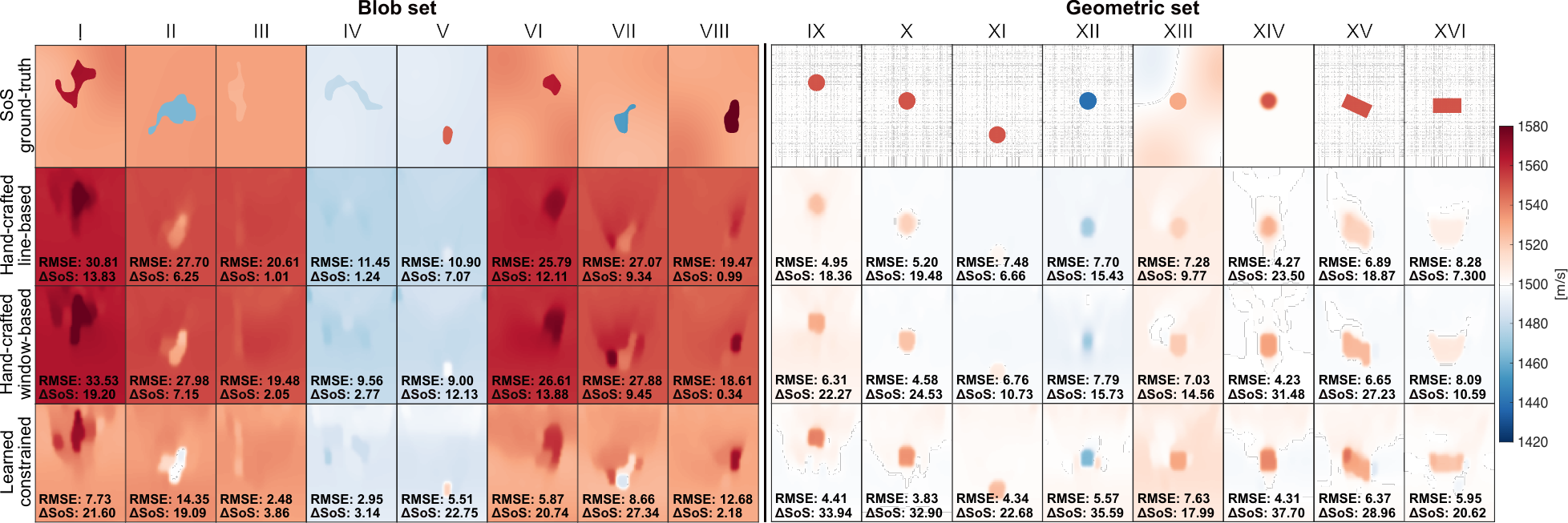}
\caption{Local SoS reconstructions for sample numerical phantoms from the Blob and Geometric sets, with varying inclusion sizes, shapes, locations, SoS contrasts and background variation.
The proposed learned models is seen to yield superior reconstructions. 
}
\label{fig:all_recons_circle_blob}
\end{figure*}

For the kernels learned above using Blob set training samples, we evaluated the performance on Blob set test samples as well as the Geometric set, as shown in \cref{fig:DTRMSE_RMSE_DeltaSoS}.
As seen, the window based model is superior to the line based model, while our learned kernel approach is superior to both hand-crafted models. 
Specifically, in Blob set the median RMSE$_t$ and RMSE$_c$ are reduced respectively by 34.2\% and 38.2\% while the median $\Delta$SoS is increased by 320.9\% compared to the conventional hand-crafted line model used in earlier works. 
Similarly in Geometric set, our method achieves 34.4\% improvement in RMSE$_t$, 18.2\% improvement in RMSE$_c$, and 63\% improvement in median $\Delta$SoS in comparison to the conventional hand-crafted line-based SoS imaging model.
Note that all the methods perform better in Geometric set compared to Blob set, since the latter contains challenging inclusion shapes with small geometric features and large SoS variations. 
To demonstrate improvements in individual local SoS reconstructions, eight samples from each test set are exemplified in \cref{fig:all_recons_circle_blob}. 

\subsection{SoS Reconstructions: Tissue-Mimicking Phantom}

\begin{figure}
\centering
\includegraphics[width=\linewidth]{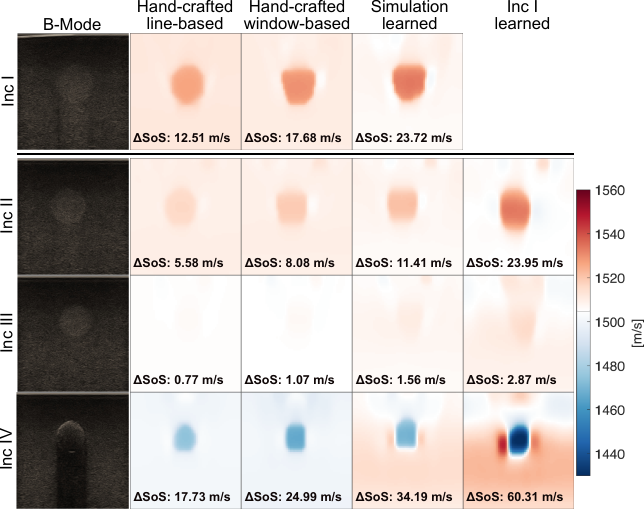}
\caption{SoS reconstructions with data acquired from four phantom inclusions in rows.
Columns show the machine-beamformed B-mode image and the reconstructions with hand-crafted and learned models.
Inc\,I learned model is not applied for the reconstruction of this training data, and hence the missing image on top right.
Contrast values are calculated based on masks annotated on the B-Mode images. }
\label{fig:phantom_sim_kernel}
\end{figure}

We next reconstruct from the phantom data using the above three imaging models, \ie the line-based model, window-based, and our kernel learned from the simulations.
Note that the generalization of a simulation-learned model to real-life acquired data is often quite challenging.
Nevertheless, the first three columns of \cref{fig:phantom_sim_kernel} demonstrate that a simulation-learned model can perform substantially superior to the hand-crafted alternatives even on machine-acquired actual data.
Compared to the conventional line-based model, our simulation-learned model doubles the SoS contrast for all inclusions.

We further investigated whether machine-transducer specific models can be learned.
Since we only had four inclusions, we did not have many independent samples to jointly learn from.
Thus we chose a single image (Inc\,I) to learn the imaging model, which exemplifies a one-shot learning case with extreme data scarcity.
We used the delineated Inc\,I mask with the manufacturer-reported SoS values as the ground-truth $\boldsymbol{s}$, and the time-delay data $\boldsymbol{t}$ from our acquisition and processing pipeline described earlier. 
This Inc\,I-learned model is then used for testing on the other inclusions, \ie to reconstruct SoS maps from their acquired data.
Using the model learned machine-specifically doubles the SoS contrast by another fold over the simulation-learned model, thus remarkably quadrupling the contrast compared to the conventional line-based model.

\subsection{SoS Reconstruction: In-vivo}
\begin{figure}
\centering
\includegraphics[width=\linewidth]{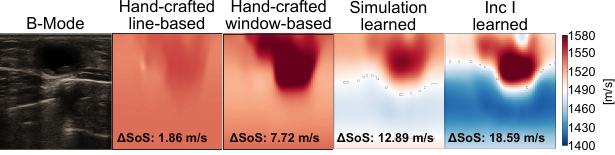}
\caption{SoS reconstructions of a clinical breast data from a patient with biopsy-confirmed ductal carcinoma. 
Contrast values are calculated using the masks annotated from B-Mode images. }
\label{fig:invivo_results}
\end{figure}

\cref{fig:invivo_results} demonstrates the SoS reconstructions of clinical breast data from a patient with biopsy-confirmed ductal carcinoma, using conventional line-based, our extension window-based, and our proposed learned model both from the simulations and from the phantom data. 
With the line-based hand-crafted model, the SoS contrast is very low with the cancerous inclusion being barely visible (without colorbar manipulation) in the reconstruction. 
The hand-crafted window model improves the contrast, while the simulation-learned model provides further improvement.
With the imaging model learned machine-specifically enhances the contrast by a large extent comparatively.

\section{Discussion and Conclusion}

In this work, we have proposed a novel method for learning the forward model of speckle-shift based pulse-echo SoS imaging. 
The accuracy of analytical methods~\cite{jaeger_computed_2015,sanabria_spatial_18} heavily depends on the imaging model adequately representing the underlying physical and numerical processes.
Variational-network based deep-learning methods~\cite{vishnevskiy_image_2018, melanie_training_2020} use such imaging models for training as well as an internal constant network operation, therefore these methods are also strongly dependent on the imaging model accuracy. 
Besides local SoS reconstruction, similar imaging models encoding tracking-SoS sensitivities are also required in other methods, \eg for the analytical estimation of global beamforming SoS in~\cite{analytical_bezek_2023}, which may benefit from our learned modeling method.

Our approach reformulates the SoS imaging problem in a convolutional form that reduces the degrees of freedom required to represent the model, and then learns the convolution kernel representing such model. 
This exploits the imaging problem structure to alleviate the complexity associated with learning the entire forward model matrix. 
To further stabilize such learning, beam widening is included as a constraint, which is then learned as a windowing function along the wave propagation path. 
This further reduces the degrees of freedom learned and hence computational complexity, also improving the problem conditioning. 

We have extensively evaluated our method on data from numerical simulations, tissue-mimicking phantom, and an in-vivo patient. 
The results show that our method produces imaging models that better fit the observed displacements and that improve SoS reconstruction accuracy and contrast compared to hand-crafted approaches. 
Our method obviates the manual effort of devising and designing complex physics-based SoS imaging models, while also implicitly incorporating effects that cannot be easily anticipated.
For example, potential artifacts originating from suboptimal beamforming, displacement tracking, or other numerical operations can be implicitly and auto-magically incorporated into such learned model.

Our method demonstrates the learning of machine-specific imaging models, even from a single image acquisition.
These are shown to be superior to simulation-learned kernels and to successfully generalize to acquired data even in in-vivo settings.
Learning machine-specific models, however, require imaging data of known SoS distributions, such as a tissue-mimicking phantom. 
We utilized the manufacturer-reported SoS values as ground-truth, which can also be independently measured as they may vary due to, \eg storage conditions, drying, or temperature variations. 

The preliminary results in clinical breast data highlight the generalizability of the learned kernels and the potential clinical applicability of our approach.
Since we synthetically create plane-wave transmissions from single-element data, any unavoidable motion present during in-vivo acquisition might reduce the signal-to-noise ratio of such synthetic PW data~\cite{schweizer_robust_23}. 
Physical PW Tx acquisitions could thus potentially improve the in-vivo results.
Nevertheless, we show that both our simulation- and phantom-learned kernels that result from motion-free data still perform quite remarkably on in-vivo data with inherent potential motion.
Note that using our presented methods, others with different machines, transducer geometry, acquisition parameters, Tx pair combinations, and so on can learn imaging models specific to their settings.

Our convolutional solution assumes a shift-invariant kernel operator over the displacement tracking field, which is true for the PW Tx employed.
For diverging wave Tx~\cite{rau_SoS_2019}, however, the paths in the imaging model vary over the displacement tracking grid points and thus the presented convolutional method does not directly apply in that setting.
The methods proposed herein for the SoS imaging problem may, nerertheless, be applicable in other imaging problems of similar nature where the forward problem adheres to a convolutional structure.
Furthermore, our constrained approach may find utility in scenarios where a forward problem operator can be dissected into multiple constituents with a mix of predictable and varying nature.

\ifCLASSOPTIONcaptionsoff
  \newpage
\fi

\bibliographystyle{IEEEtran}
\bibliography{IEEEabrv,bib/cite}

\end{document}